\documentclass[aps,pra,twocolumn,showpacs,groupedaddress,letterpaper,nofootinbib]{revtex4-1}

\usepackage{graphicx}
\usepackage{mdframed}
\usepackage{framed}
\usepackage{indentfirst}
\usepackage{natbib}
\usepackage{amsmath,amssymb}
\usepackage{subfigure}
\usepackage{enumitem}
\usepackage{mdframed}

\usepackage{url}
\usepackage{breakurl}
\usepackage[breaklinks]{hyperref}

\usepackage[nodisplayskipstretch]{setspace}

\newcommand{\bw}{\textcolor{black}}
\newcommand{\qr}{\textcolor{black}}

\begin{document}

\title{Student Difficulties with Boundary Conditions in Electrodynamics}


\author{Qing X. Ryan}
\affiliation{Department of Physics and Astronomy, California State Polytechnic University, Pomona, California, 91768}

\author{Bethany R. Wilcox}
\affiliation{Department of Physics, University of Colorado, 390 UCB, Boulder, Colorado, 80309}

\author{Steven J. Pollock}
\affiliation{Department of Physics, University of Colorado, 390 UCB, Boulder, Colorado, 80309}

\begin{abstract}
Identifying and understanding student difficulties with physics content in a wide variety of topical areas is an active research area within the PER community. In many cases, physics topics appear multiple times in different contexts across the undergraduate physics curriculum.  As these common topics reappear, students' difficulties can perpetuate from one context to the next, or new difficulties can emerge as students encounter new physical contexts. One example of such a topic is boundary conditions. It is broadly considered to be an important topic that advanced physics undergraduates are expected to understand and apply in multiple topical areas including classical mechanics, quantum mechanics, and throughout electricity and magnetism (E\&M). We report findings from an investigation of student difficulties using boundary conditions focusing on the context of electrodynamics. Our data sources include student responses to traditional exam questions, conceptual survey questions, and think-aloud interviews. The analysis was guided by an analytical framework that characterizes how students activate, construct, execute, and reflect on boundary conditions. Common observed student difficulties include: activating boundary conditions in appropriate contexts; constructing a complex expression for the E\&M waves; mathematically simplifying and manipulating complex exponentials and, checking if the reflection and transmission coefficient are physical. We also present potential pedagogical implications based on our observations.
\end{abstract}
\maketitle

\section{\label{sec:intro}Introduction}

Identifying and understanding student difficulties with physics content in a wide variety of topical areas has been, and continues to be, an active research area within the PER community \cite{Meltzer2012}. Much of this work has focused on students and topics in introductory physics; however, a growing body of research suggests that upper-division students continue to struggle with problem solving even in the advanced physics courses \cite{caballero2015, pepper2012}. Student difficulties in upper-division problem solving stem in part from the more complicated math and more sophisticated physics characteristic of upper-division content.  However, difficulties in the upper-division can also stem from the cyclic nature of the physics curriculum, in which some physics topics appear several times in different contexts across the undergraduate physics curriculum \cite{Manogue2001}.  For these topics, difficulties that are not addressed in early courses can persist and become worse when the idea appears again in a more advanced course.  

One idea which appears several times throughout the undergraduate curriculum is that of boundary conditions (hereafter BCs). In physics, BCs generally refer to the conditions physical quantities must satisfy at the boundary between two regions.  BC problems are a subset of boundary value problems more generally, which also include problems that require applying initial conditions (i.e., boundary conditions in time) \cite{mathphysics}.  BCs, and boundary value problems more generally, are ubiquitous in both physics and mathematics, for example, boundary value problems always appear in the context of differential equations \cite{Gladwell2008}.   However, BCs are particularly critical in physics because they are often necessary in order to change general and abstract mathematical expressions into physically meaningful solutions which have descriptive and predictive power within a particular physical system \cite{mathphysics}. BCs are used in a variety of topical areas in physics including classical mechanics, quantum mechanics, and throughout electricity and magnetism (E\&M).  For example, BCs are used when analyzing waves on a finite string in classical mechanics \cite{classicalmechanics}.  BCs also appear four separate times in a canonical junior-level E\&M in text (Griffiths chapter 3, 6, 7 and 9) \cite{Griffiths}.  The importance of BCs in quantum mechanics has been pointed out \cite{madrid2003}, for example BCs are important in solving tunneling problems.

There are relatively few existing research studies investigating student difficulties with BCs. Wilcox \emph{et al.} \cite{Wilcox2015} looked at student difficulties with separation of variables in the context of electrostatics, where applying boundary conditions is an essential part of the problem solving process. They found that, particularly in the case of spherical separation of variables, some students struggled to identify and express the appropriate BC when it was not explicitly provided in the problem prompt.  However, the results from Wilcox \emph{et al.} were limited to students' difficulties with BCs for the scalar potential and did not address students' difficulties with BCs with respect to vector quantities.  McKagan \emph{et al.} \cite{mckagan2008} reported their findings on student difficulties with tunneling in quantum mechanics. While applying BCs can be an essential part to solve tunneling problems, their main focus was mostly the conceptual aspect, while we focus on the problem solving process of applying BCs (in E\&M). The findings presented here have some interesting overlap with those of McKagan et al.  However, the different physical contexts of these works (i.e., E \&M vs. Quantum mechanics) also resulted in some notable differences in our findings. We will discuss the similarities and differences with our findings in more detail in Sec. \ref{sec:tunneling}.

Here, we contribute to the body of knowledge around student understanding of BCs by investigating student difficulties in another context: electrodynamics.  In particular, we focus on BCs as applied to solving Maxwell's equations for electromagnetic (EM) waves incident on a boundary between two linear media.  In this situation, Maxwell equations result in four equations (Eqns.\ \ref{eq:eqn1}-\ref{eq:eqn4}) that the electric and magnetic fields must satisfy at a boundary. An example of the derivation of these equations is presented in chapter 7 in Griffiths' book \cite{Griffiths}. 
\begin{gather}\label{eq:eqn1}
\vec E^\parallel_1=\vec E^\parallel_2
\end{gather}
\begin{gather}\label{eq:eqn2}
\epsilon_1E^\perp_1- \epsilon_2E^\perp_2= \sigma_f
\end{gather}
\begin{equation}\label{eq:eqn3}
B^\perp_1=B^\perp_2
\end{equation}
\begin{equation}\label{eq:eqn4}
\vec B^\parallel_1 / \mu_1-  \vec B^\parallel_2 / \mu_2= \vec K_f \times \hat{n}
\setstretch{0.1}
\end{equation}
In the case of no free charge or current at the boundary, Eqn.\ \ref{eq:eqn2} and Eqn.\ \ref{eq:eqn4} become: 
\begin{equation}\label{eq:eqn5}
\epsilon_1E^\perp_1= \epsilon_2E^\perp_2
\end{equation}
\begin{equation}\label{eq:eqn6}
\vec B^\parallel_1 / \mu_1=  \vec B^\parallel_2 / \mu_2
\end{equation}

\noindent Eqn.\ \ref{eq:eqn5} and Eqn.\ \ref{eq:eqn6} are the equations more commonly used in the \bw{canonical} BCs problems that appeared in homeworks and exam questions investigated by this study.

In this paper, we utilize an existing analytical framework\cite{Wilcox2013, Wilcoxdeltafunction} to structure our investigation of student difficulties with the use of BCs in electrodynamics. \bw{This framework is discussed in more detail in the following section (Sec.\ \ref{sec:acer}}).  We summarize common student difficulties identified with our student population (Sec.\ \ref{sec:results}). We then \bw{discuss how our results align with related studies of BCs in quantum mechanics (Sec.\ \ref{sec:tunneling}), and end with} some implications for teaching, the limitations of the study and suggestions for future work (Sec. \ref{sec:conclusionsanddiscussion}).

\section{\label{sec:acer}Analytical Framework}

\bw{To guide and structure our analysis of student difficulties with BCs in the context of EM waves, we utilize an analytical framework known as ACER (activation, construction, execution, reflection) \cite{Wilcox2013}.  }The ACER framework is an analytical tool that \bw{was developed to} characterize student difficulties with upper-division problem solving by organizing the problem-solving process into four general components: \textit{Activation} of the tools, \textit{Construction} of the models, \textit{Execution} of the mathematics, and \textit{Reflection} on the results. These components appear consistently in expert problem solving \cite{Wilcox2014,Wilcox2013} and are explicitly based on a resources view on the nature of learning \cite{Hammer2000, Wilcox2013}. Since the particulars of using mathematical and physical tools to solve upper-division physics problems are highly context-dependent, ACER is designed to be operationalized for specific physics topics. The operationalization of the ACER framework for BCs will be discussed in Section.\ \ref{sec:acer oper}, and additional details about the ACER framework can be found in Ref. \cite{Wilcox2013}.

ACER was \bw{originally} designed to characterize students' difficulties using mathematical tools in physics.  \bw{The use of BCs, however, would more appropriately be characterized as the application of a physics concept/principle rather than the use of a mathematical tool. Thus,} this is a slight extension on the original use of ACER. However, \bw{the overarching goal of the ACER framework was to assist in the process of investigating mathematically sophisticated physics problem solving \cite{Wilcox2013}.  Thus, since we have found ACER to be a productive analytical tool when thinking about students' difficulties with BCs, and the theoretical grounding of the framework does not preclude the use of a conceptual (rather than mathematical) tool, we have opted to utilize the framework here.  }

\bw{Several other frameworks have been applied to investigations of} student difficulties. For example, the framework of epistemic games presented by Tuminaro \cite{Tuminaro} was developed to analyze students' problem solving behavior in terms of locally coherent goal-oriented activities. These games both guide and limit what knowledge students think is appropriate to apply at a given time. Identifying these games provides a way of parsing students' tacit expectations about how to approach solving physics problems \cite{Tuminaro2007}. This framework was developed for introductory students and relies on students' explicit discussion of the details of their work. Another framework developed by Bing to analyze upper-level students' use of mathematics was epistemic framing \cite{Bing}. Epistemic framing  is the students' unconscious answer to the question: ``What kind of activity is this?'' A student's framing can be identified by examining the justifications and proof students offer to support their mathematical claims \cite{Bing}. Earlier research on investigating students' understanding includes diSessa's work on epistemology in physics \cite{diSessa1993, diSessa1998}, Sherin's work on symbolic forms \cite{Sherin1996,Sherin2001}, and Collins \emph{et.al}'s work on Epistemic forms and epistemic games \cite{Collins1993}. 


\bw{One significant difference between these theoretical frameworks and ACER is that these frameworks usually require detailed analysis of students' in-the-moment reasoning through video or audio} data, while ACER can be used to analyze written solutions. Connections between these, and other, theoretical frameworks and ACER is discussed elsewhere \cite{wilcox2015thesis}.  For the purpose of this study, we are interested in identifying the points (nodes) in students' work where key difficulties appear. \bw{ACER provides a mechanism to identify these nodes using students' written solutions.  The framework} also informed the development of interview protocols that targeted \bw{more aspects of the problem-solving process, including those} typically bypassed by regular exam questions. We developed interview questions to study these common difficulties in more depth (see Sec.\ \ref{sec:datasource}).

\section{\label{sec:datasource}Data Sources}



Data for this study were collected from the context of the Junior level electrodynamics course at the University of Colorado Boulder (CU). The electrodynamics course at CU (E\&M II: Griffiths \cite{Griffiths} Ch.7-12) is the second semester of the electricity and magnetism sequence. The student population is composed of physics, astrophysics, and engineering physics majors with a typical class size of 30-60 students. At CU, E\&M II is often taught with varying degrees of active engagement through the use of research-based teaching practices \cite{Chasteen2012}, such as peer instruction \cite{Mazur} and in-class tutorials \cite{Chasteen2012}. 

\begin{figure}
\includegraphics[scale=0.7]{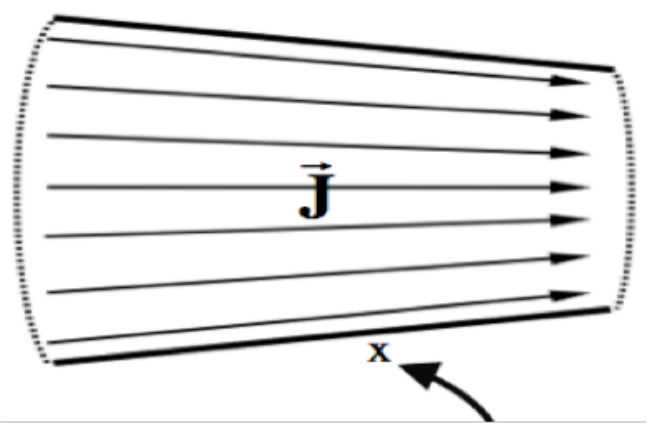}
\caption{Question on the CURrENT: A steady current flows in a long wire that has a uniform conductivity $\sigma$.  Is the electric field \underline{just} \underline{outside} the surface of the wire (e.g., at the point ``x" shown in the diagram) zero or non-zero?}\label{fig:currentquestion}
\end{figure}

\qr{There are three major data sources for this study. First, students' solutions to traditional midterm exams (three semesters) from CU (N=158) \footnote{Since we are interested in the frequency of certain mistakes, for exam data, we are reporting the number of solutions instead of the number of students.} Of the exam solutions,128 were to a question that targeted \textit{Construction} and \textit{Execution} of the ACER framework and 30 were to another question that targeted \textit{Reflection}. The second data source was responses to one question (Fig.\ \ref{fig:currentquestion}) on the \textit{Colorado UppeR-division ElectrodyNamics Test} (CURrENT: a conceptual assessment given to many students in electrodynamics) \cite{Baily2012,Ryan2014} (N=500 from 9 institutions). The final data sources were two sets of think-aloud interviews (N=11) with CU students.}

The two semesters of exam data collected for this study were taught by two different instructors, one was a PER (physics education research) faculty member who is also one of the authors of this paper (S.J.P), the other was a traditional physicist but had used some of the materials developed by the PER group such as clicker questions, in-class tutorials and homework/exam problems. These exam questions analyzed below all posed a situation involving electromagnetic waves at the interface of two media and require using Boundary conditions. \qr{An example of such a situation is presented in (Fig.\ \ref{fig:BCprob})}. Exams were analyzed by coding common student ideas and/or difficulties that appeared at each element of the operationalized framework. Exam questions mostly targeted Construction and Execution components, and did not provide a complete picture of the Activation or Reflection components. To address all aspects of the ACER framework, we relied further on data from CURrENT and interviews. 

The CURrENT \cite{Baily2012} is a research-based assessment that was developed to assess students' conceptual understanding of electrodynamics. This assessment can be accessed at \cite{perwebsite}. Validity and reliability studies for this assessment have been conducted \cite{Ryan2014}. One question asks if the $\vec {E}$ field just outside of a wire carrying steady current is zero or non-zero (Fig.\ \ref{fig:currentquestion}) and also asks students to provide reasoning. According to BCs (equation 1), the $\vec {E}$ field just outside of the wire must be non-zero. The total number of solutions for this data set is 500 with 38\% from CU students (N=191), 49\% are from 3 other Ph.D granting (in Physics) institutions (N=242), and 13\% are from 5 other Bachelor’s degree (in Physics) granting institutions (N=67). 


Think-aloud interviews were conducted in two sets, one performed about six months after the other.  Interviewees were paid volunteers who completed the E\&M2 class one or two semesters earlier and responded to an email request for research participants. All interviewees took the course during one of the semesters for which we have exam data. Final course grades for interview participants with mostly A and B students (6 A \& A- students,  4 B students and 1 C- student). Interview protocols for both sets of interviews were designed, in part, to probe more deeply into difficulties identified in the exam solutions, and thus included one or two exam-style questions. In addition, the interview protocols also included questions designed to target aspects of the ACER framework not accessed by the exams. These questions will be discussed in greater detail later in the paper.



\section{\label{sec:acer oper}ACER Operationalization}

\begin{figure}
\includegraphics[scale=0.5]{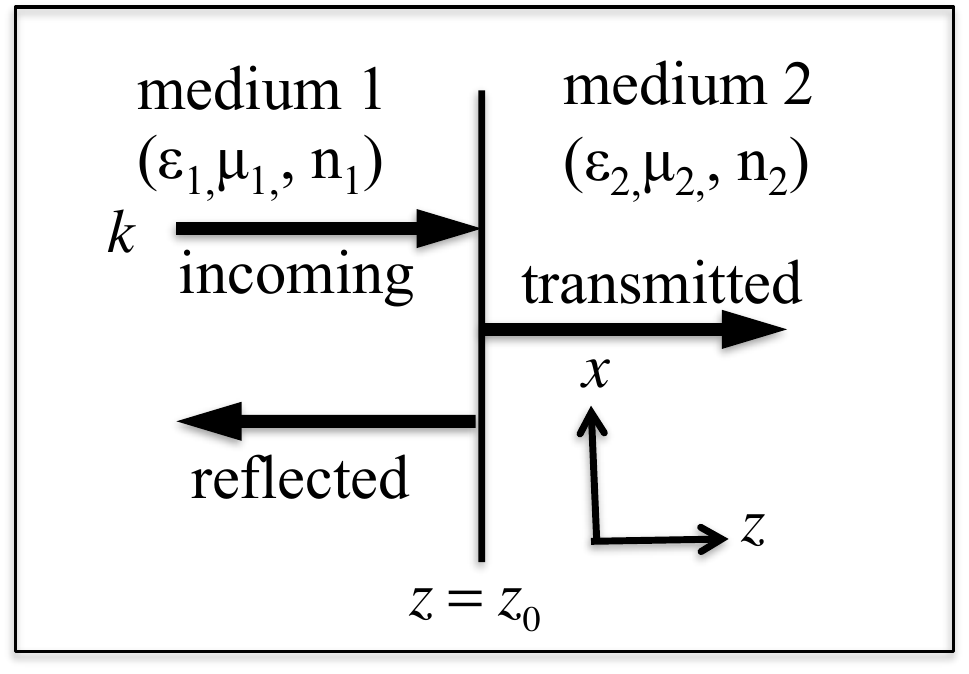}
\caption{Setup of a typical BCs problem in the context of electromagnetic waves. Note that exam questions included both normal and oblique incidence situations.}\label{fig:BCprob}
\end{figure}

We operationalized ACER for the use of BCs primarily in solving electromagnetic wave problems (Fig.\ \ref{fig:BCprob}). The operationalization process involves a content expert working through the problems of interest while carefully documenting their steps and mapping these steps onto the general components of the framework. The resulting outline is discussed with other content experts and refined until consensus is reached that the key elements of the problem have been accounted for. Further refinement are made (if necessary) to accomodate unanticipated aspects of student problem solving when applying this coding scheme to students' work. A skeletal summary of the operationalized framework is listed below. A detailed description of the operationalization process is discussed elsewhere \cite{Wilcox2013}. 

{\bf Activation of the tool}  involves identifying BCs as a relevant physical tool. BCs are likely to be activated in two situations:
\vspace{-2mm}
\begin{enumerate}[label=A\arabic*:, align=left] \itemsep-2pt
  \item Use BCs when being explicitly asked to (this would be considered as bypassing Activation).
  \item Use BCs when being presented a physical situation involving two media and asked for the relationship between physical quantities across the boundary.
\end{enumerate}
\vspace{-2mm}

{\bf Construction of the model}  involves mapping boundary conditions onto the specific physical situation/system. We operationalized Construction into three elements. The numbering of these elements is only for labeling purposes and does not necessarily indicate the order of the problem solving process. 
\vspace{-2mm}
\begin{enumerate}[label=C\arabic*:, align=left] \itemsep-2pt
  \item Write mathematical expressions of the complex fields for the incoming, reflected, and transmitted waves (hereafter denoted by the subscripts: in, refl, and trans).
  \item Select the appropriate BCs to be used with the corresponding components of the fields, which involves knowing the geometry of the situation (e.g. which component of the fields is the parallel or perpendicular component).
  \item Set up equations at the boundary which includes superposing the fields of the incoming and reflected waves, and applying the equations only at the boundary.
\end{enumerate}
\vspace{-2mm}

{\bf Execution of the mathematics} involves performing mathematical steps to reduce the result to a form that can be interpreted. 
\vspace{-2mm}
\begin{enumerate}[label=E\arabic*:, align=left] \itemsep-2pt
   \item Further simplify the expression after setting up up at the boundary, including regrouping or cancelling exponential terms. (\qr{Note that the use of complex exponentials here could be its own full investigation.  However, identifying student difficulties with complex exponentials is not the focus of this study and therefore we did not sub-operationalize the ACER framework for the use of complex exponentials by itself.})

   \item Calculate the Reflection and Transmission coefficient using the amplitudes of the E field. This element was not analyzed in this study because none of the questions included in this study elicited or required this element.
\end{enumerate}
\vspace{-2mm}

{\bf Reflection on the results} involves reflecting on, and evaluating the mathematical results to gain physical insight and ensure consistency --- a practice common among content experts. There are two common ways to reflect on BCs problems: 
\vspace{-2mm}
\begin{enumerate}[label=R\arabic*:, align=left] \itemsep-2pt
  \item Check the units of the relevant quantities.
  \item Check limiting behaviors. (e.g., the reflection coefficient R and transmission coefficient T must satisfy these conditions: $0 \leq R \leq 1$, $0 \leq T \leq 1$, $R+T=1$).
\end{enumerate}
\vspace{-2mm}

\section{\label{sec:results}RESULTS}
Using the operationalized ACER framework, we identified several common student difficulties. To organize the presentation of these difficulties, we group them according to the ACER components.

\subsection{\label{sec:activation}Activation of the tool}
Most exam questions in the context of electromagnetic (E\&M) waves bypass Activation by giving an explicit prompt (A1). Not only do these questions tell students to use BCs but also list the equations of BCs (Eqns.\ \ref{eq:eqn1}, \ref{eq:eqn3}, \ref{eq:eqn5} \&\ref{eq:eqn6}) explicitly for students. Therefore, in think-aloud interviews we investigated Activation by observing whether students were able to activate BCs even when they were not explicitly provided.

In the interviews, we gave an E\&M wave problem to three students and asked them to use BCs but did not provide the equations (Eqns.\ \ref{eq:eqn1} - \ref{eq:eqn6}).  The question provides a figure similar to Fig.\ \ref{fig:BCprob} and asks students to write down a mathematical expression that relates the amplitude of the electric fields of the incident, reflected and transmitted waves. Successful activation of BCs involves writing down $\vec E^\parallel_1=\vec E^\parallel_2$ (Eqns.\ \ref{eq:eqn1}) or any other equations (Eqns.\ \ref{eq:eqn2}-\ref{eq:eqn6}). One student was able to derive the BCs from Maxwell’s equations. Two students did not write down the BCs but rather, jumped straight to this result (with a sign error on $\tilde{E}_{0R}$): $\tilde {E}_{0I}=\tilde {E}_{0R}+\tilde {E}_{0T}$ (tilde means a complex quantity). The reasoning they provided to justify their answer looked like this: ``\textit{You can't just come away with more waves than you had coming in. Some of it will go through, some of it will go back. It has to add up, kind of like the conservation}." Even being asked explicitly to use BCs, when these equations are not provided, their reasoning showed that they were not using BCs. \qr{One might argue that this is still an activation of a different kind of boundary condition, for example, an idea of conservation of energy. However, it is not the correct activation for BCs for $\vec{E}$ and $\vec{B}$} that are necessary to solve this problem.

In addition to investigating student difficulties with Activation in the context of E\&M waves, we expanded our investigation by shifting to a different context. One question on the conceptual test (CURrENT \cite{Ryan2014}) asks if the $\vec {E}$ field just outside of a wire carrying steady current is zero or non-zero (Fig.\ \ref{fig:currentquestion}). It also asks students to provide reasoning. This falls into ACER category A2. According to BCs (Eqns.\ \ref{eq:eqn1}), the $\vec {E}$ field just outside of the wire must be non-zero. 155 students (31\%) answered this question correctly both for the correctness part (E is non-zero) and the reasoning part (either by invoking BCs or stating there are surface charges). By coding all student responses, we found that only 26\% of the students (130 out of 500) explicitly used BCs (students who activated BCs mostly also correctly conclude that E is non-zero), the rest of the students failed to activate BCs on their own in this context (i.e. did not use BCs in their resoning). 

\begin{figure}
\includegraphics[scale=0.5]{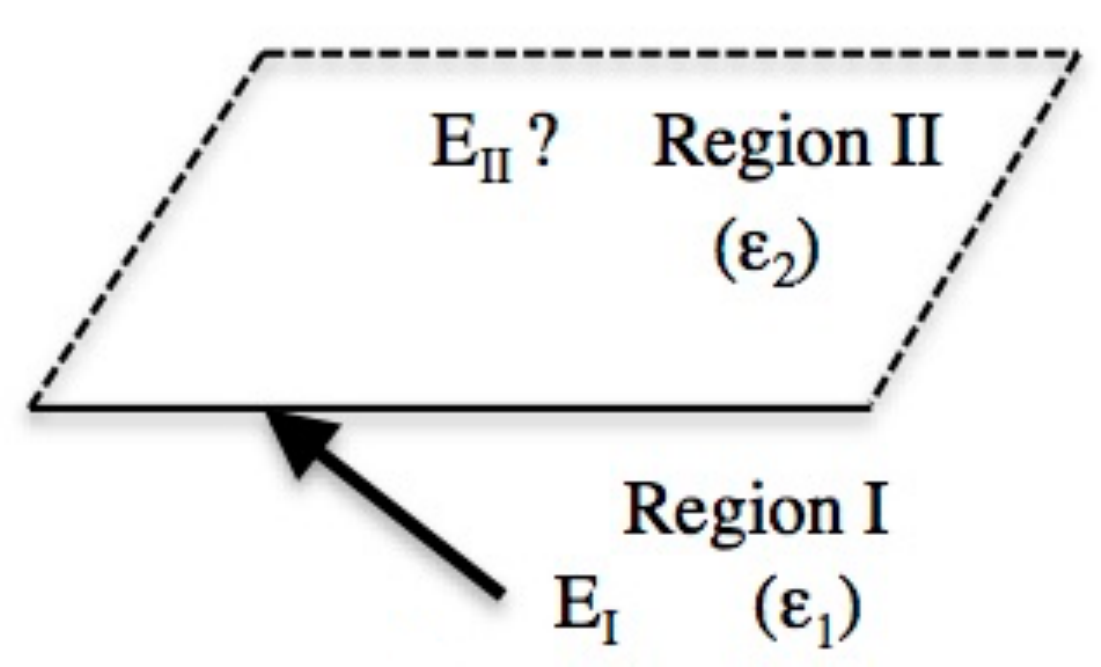}
\caption{Scaffolding question: $ \epsilon_1= 2\epsilon_2$, $E_I$ in region 1 was shown in the picture. Use boundary conditions ($\vec E^\parallel_1=\vec E^\parallel_2$, $ \epsilon_1E^\perp_1= \epsilon_2E^\perp_2$) to draw an arrow that represents $E_{II}$ on the other side.}\label{fig:scaffoldquestion}
\end{figure}

We further explored the nature of student difficulties with the CURrENT question by conducting think-aloud interviews (N=6). Different levels of cues were provided by the interviewer (cues were provided to 5 out of the 6 students): a) providing the Griffith's book\cite{Griffiths} and asked students to review the chapter on electromagnetic waves before the interview (1 student); b) providing a scaffolding question (Fig.\ \ref{fig:scaffoldquestion}) before giving the CURrENT question (3 students); c) asking students to redo this question once they solved a E\&M wave problem where BCs were activated (1 student).  

Only two interviewees who received cues a and b respectively activated BCs. The remaining students were asked by the interviewer explicitly to use BCs once they failed to activate on their own. Even after being asked to use BCs, two students still didn't think that BCs were applicable. One student said: ``\textit{but that is light, this is not light}". Another student used BCs after being asked and correctly indicated $\vec {E}$ is nonzero just outside of the wire but still was not satisfied with the answer: ``\textit{my answer is correct if the boundary conditions are true. I don't know if these boundary conditions were universal or only work for specific conditions.}" So even after explicit or implicit cues from the interviewer, those students still had trouble activating BCs. 

In summary, we found that, in the absence of direct activation (A1), many students failed to activate BCs on their own. In the context of E\&M waves, some students still had trouble activating the correct BCs equations when they were not explicitly provided. When shifting to a different context (wire with steady current), only two (out of six) students activated BCs while the remaining students were unsure if BCs were applicable because it was not a light or wave context, even after being explicitly prompted.

\subsection{\label{sec:construction}Construction of the model}
When constructing a complex expression for the $\vec {E}$ and/or $\vec {B}$ field (element C1)(e.g.Eqn.\ \ref{eq:eqn7}, it is important to know the correct direction for the wave vector $\vec {k}$ and fields $\vec {E}$ and/or $\vec {B}$. 


\begin{equation}\label{eq:eqn7}
\vec {\tilde{E}}_{in}=\tilde{E}_{0,in}exp [i(\vec{k} \cdot \vec{r}-\omega t)] \hat{x}
\end{equation}

$\vec{k}$ is the wave vector, $\vec{r}$ is the position vector, $\omega$ is the angular frequency, t is time. In the case of the reflected wave, the direction of $\vec{E}_{refl}$ depends on the index of refraction of the two media. In the case of normal incidence such as Fig.\ \ref{fig:BCprob}, when $n_1 > n_2$ , the direction of $\vec{E}_{refl}$ is the same as $\vec{E}_{in}$. When $n_1 < n_2$,  $\vec{E}_{refl}$ is out of phase with $\vec{E}_{in}$ which means $\vec{E}_{refl}$ will be in the opposite direction as $\vec{E}_{in}$. When asked explicitly to draw arrows representing the direction of the fields, a third of the students (31\%, 15 out of 48) incorrectly flipped the direction of $\vec{E}_{refl}$ in the case of $n_1 > n_2$. The total number of solutions here is less than 128 because we only count the problems that include this construction component, not all problems include all ACER elements. The same reason applies for other numbers (less than 128) reported later in the paper.  It is impossible to interprete from the written solutions why students think $\vec{E}_{refl}$ should be flipped, unless they happened to explicitly articulate their reasoning in writing and most of the students did not. One possible hypothesis is that since the reflected wave is propagating in the opposite direction, students believe the direction of the $\vec{E}$ field is flipped as well. We did not observe evidence from the written work that students were making connections between the direction of $\vec{E}_{refl}$ and the relationship of the index of refraction between the two media.

Other errors commonly occurred when simplifying $\vec {k} \cdot \vec{r}$ ($\vec{k}$ is the wave vector, $\vec{r}$ is the position vector). In the case of oblique incidence, the dot product becomes $k_xx+k_yy+k_zz$. Half of the solutions (18 out of 36) did not write down this expression for the dot product $\vec {k } \cdot \vec{r}$ first. \qr{Writing down this dot product explicitly suggests further expressions of the complex notations of $\vec {E}$ and/or $\vec {B}$ field were based on the calculation of the dot product between two vector quantities. These students (18 out of 36) treated this as a simple product between k and r, or x (or y, z) as if they were scalar quantities.} Out of those who calculated the dot product, only 4 worked it out correctly. Various mistakes were made in the rest of the solutions: replacing $\vec {k} \cdot \vec{r}$ with $kz$, $kr$, $krsin\theta$ or $krcos\theta$. In the case of normal incidence where $\vec{k}$ is propagating in the z direction (Fig.\ \ref{fig:BCprob}), this dot product for the reflected and transmitted wave simplifies into ${-k_{refl}z}$ and $k_{trans}z$ with $k_{refl}$ and $k_{trans}$ being the magnitude of the wave vector in the two media. Working out the dot product for normal incidence is much easier. However, almost a fifth (18\%, 23 out of 128) of the solutions missed the negative sign for $k_{refl}$, and more than a tenth (13\%, 16 out of 128) of the solutions did not differentiate between $k_{refl}$  and $k_{trans}$.

For normal incidence, students also need to recognize that $\vec {E}$ is parallel to the boundary and select the corresponding BCs (element C2).  About a tenth (11\%, 14 out of 128) of the solutions chose the wrong BCs equation (tried to use the BC on the perpendicular component). In the case of oblique incidence, students need to determine the correct components of $\vec {E}$ to be used with the corresponding boundary conditions (element C2). Over a third (33\%, 12 out of 36) of the solutions had the wrong components (e.g., use $sin$ instead of $cos$). There was not enough evidence from the written solutions to tell if students got confused about what E might be parallel or perpendicular to (i.e., is it the surface, or the normal vector to the surface) or simply made a trigonometry error. \qr{However, none of the students from the interviews (N=4) with the same question made such trigonometry errors, suggesting that the latter is unlikely to be the dominant reason.}

Since there exist both incoming and reflected waves in medium 1 (Fig.\ \ref{fig:BCprob}), $\vec {E}$ in medium 1 can be written as $\vec{E}_{in}+\vec{E}_{refl}$ (element C3). This is superposition of two vector quantities and the directions need to be considered. For example, as stated earlier, if $\vec{E}_{refl}$ is the opposite direction as $\vec{E}_{in}$, then there exists a minus sign after reducing $\vec{E}_{in}+\vec{E}_{refl}$ to scalar expressions. Almost 20\% (25 out of 128) of the solutions have the wrong superposition, i.e. using a plus sign when it should be minus, and vice versa. This error is also largely dependent on whether students constructed the correct expression for $\vec{E}_{refl}$, since the direction of $\vec{E}_{refl}$ is contained in the complex expression for the fields. 

Other difficulties were also observed at this step (C3). In order to set up equations at the boundary (element C3), students need to replace $\vec {E}_{in}$  and $\vec {E}_{refl}$ with the complex expressions obtained previously (C1). However,  in over a third (36\%, 46/128) of the solutions, the $\vec {E}$ field in medium 1 was replaced with the sum of only the amplitudes of the incoming and reflected E field ($E_{0,in}+E_{0,refl}$). The exponential terms were lost in these solutions despite the fact that complex forms of the incoming and reflected $\vec {E}$ field were previously obtained. This strategy results in a correct expression only when the boundary is located at $z=0$ (Fig.\ \ref{fig:BCprob}) for normal incidence.  

We also observed that nearly half of the students (57 out of 128) omitted this step of applying the equation at the boundary. This omission was observed to happen almost always together with the previous difficulty (exponential terms were lost). Out of these 57 solutions that omitted applying the equation at the boundary, 46 of those also made the mistake of losing the exponential terms. Applying the equation at the boundary means replacing the position variable (e.g: $z$) with the specific location of the boundary(e.g.: $z_0$, $z_0$ can be zero in some problems) and the position variable is contained inside the exponential terms. It is understandable that if the exponential terms were lost, plugging in the boundary location would be lost as well.

As mentioned previously, the limitation of written solutions is that most of the time they do not provide evidence for why students make specific errors. However, the spontaneous comments offered by three students on their exam might shed some light on the reasoning. When ignoring the exponential terms and keeping only the amplitude of the E field, students arrived at the result: $|{E_{0,in}}|=|{E_{0,refl}}|+|{E_{0,trans}}|$. They justified their answer like this:``\textit{This makes sense since the incident wave gets partially reflected and partially transmitted}." This justification is similar to the conservation reasoning provided by the students interviewed (see previous section of ``Activation of the tool"). One might hypothesize that students omitted this piece of the ACER framework (C3) because of cognitive interference due to this conservation idea. See more discussion about this is in Sec. \ref{sec:tunneling}.

\begin{figure}
\includegraphics[scale=0.3]{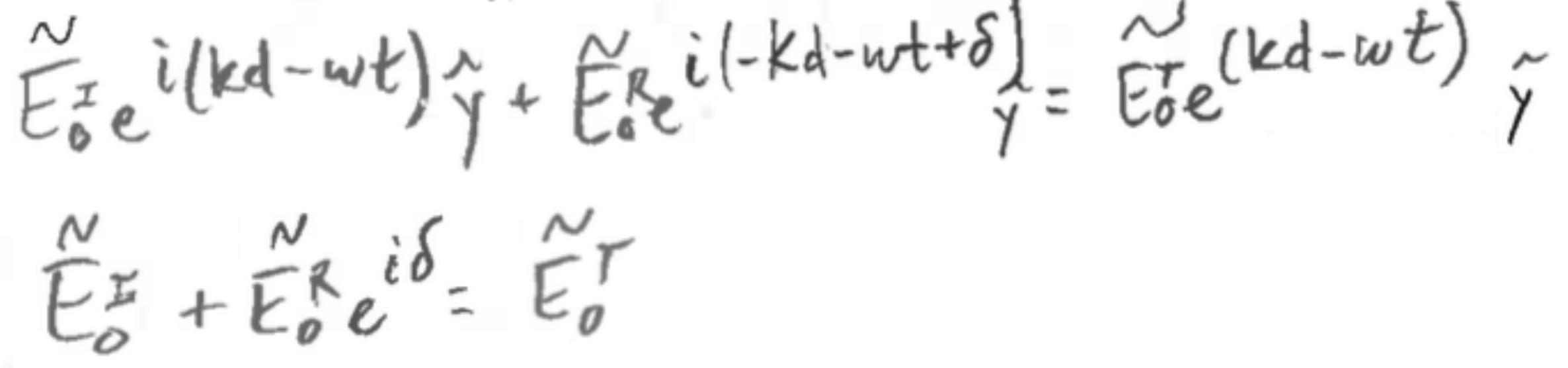}
\caption{Example of student solution where all the exponential terms were cancelled.}\label{fig:executionmistake}
\end{figure}

\subsection{\label{sec:execution}Execution of mathematics}
\begin{figure}
\includegraphics[scale=0.5]{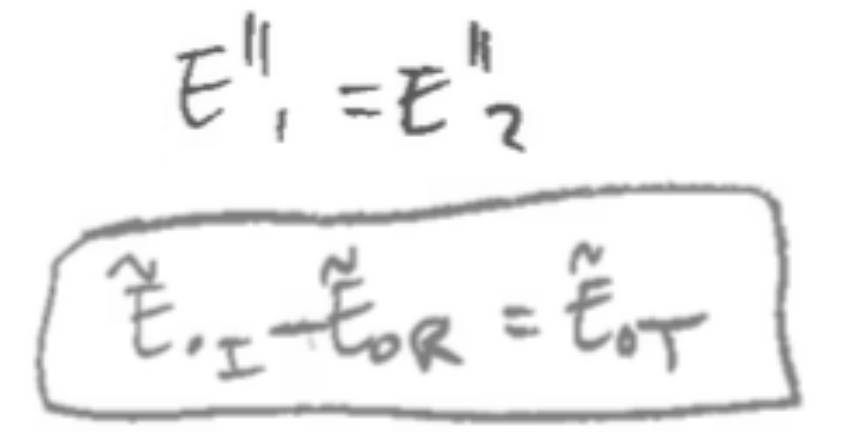}
\caption{Example student solution where the nature of the mistake is difficult to determine.}\label{fig:hardtoparse}
\end{figure}

The most common mistake observed in execution (element E1) is the inappropriate cancellation of all exponential terms. When the boundary is located at $z=0$ (Fig.\ \ref{fig:BCprob}), exponential terms go to 1, so they can be dropped. However, students still cancelled the exponentials when the boundary was located at $z=d$. Over a half of the solutions (56\%, 50 out of 90) missed the exponential terms in their final answer (See Fig.\ \ref{fig:executionmistake}), which was problematic when the boundary has a non-zero location. Parsing out the nature of this type of error can be challenging, sometimes there isn't enough evidence from the written solutions to determine whether this is a mistake about ``execution of mathematics" or ``construction of the model" in the ACER framework.  Fig.\ \ref{fig:hardtoparse} shows an example of such solutions. From the written solutions, there was often not enough evidence to tell if students forgot to carry the exponential terms into the boundary conditions (which is a construction error as described in the previous section), or if they simply cancelled the exponentials in their head without writting it out on paper (an execution error). 
Out of the 50 solutions that missed the exponential terms, almost half (23 out of 50) of the solutions did show evidence that they were indeed cancelling the exponentials  (See Fig.\ \ref{fig:executionmistake}), suggesting this is an execution error rather than a construction error.

\qr{The cancellation was coded as an Execution error in ACER; however, an execution error does not necessarily mean a pure math mistake. In other words, execution errors are math mistakes, but they may not indicate that the students are incapable of correctly executing the mathematics more generally.} In order to further investigate the nature of this issue, we conducted think-aloud interviews with a similar expression written in a pure math context. Symbols that represent physical quantities were replaced with arbitrary math variables (wave vector k, speed of light c, time t were replaced with d, a, b):
\begin{framed}
\textit{The following mathematical expression is true when}\, $x=\alpha$
\begin{equation*}
Ae^{i(d_{1}x - ab)} + Be^{i(-d_{1}x - ab)}= Ce^{i(d_{2}x - ab)}
\end{equation*}
\textit{For}\, $x=\alpha$\, \textit{can you further simplify it?}
\end{framed}
None of the interview students made the same cancellation mistake. Interestingly, 3 (out of 4) students still made the connection to the wave context spontenously: `` \textit{this is similar to the E\&M waves problems}", `` \textit{I remember you can always match the coefficients in those boundary condition problems}". These interviews indicate that students’ math execution can be affected by the physics context.

Even though students in this pure math context did not make the same cancellation mistake, they still struggled with complex exponential calculations. Two students made major math errors with complex exponentials: (e.g., $e^{A+B}=e^A+e^B$ and $e^{-A}=e^{-1}e^A$). \qr{Note that the use of complex exponentials could be its own full investigation as mentioned previously in Sec.\ \ref{sec:acer oper}. However, identifying student difficulties with complex exponentials is not the focus of this study and therefore we do not list all the mistakes on the use of complex exponentials here.}


\subsection{\label{sec:reflection}Reflection on the results}
Evaluating a solution by checking units and limiting behaviors is an important skill that physicists value. Our regular exam questions on BCs did not access Reflection explicitly and we saw almost no spontaneous reflection in the written work. Therefore we constructed an interview question to further investigate how students reflect. We presented students with incorrect but plausible expressions of the reflection and transmission coefficients R and T in a novel situation. Students were asked to examine if these expressions make sense without going through the calculation.  Three out of the 4 students interviewed were able to spontaneously point out that the T provided was not unitless (R1). The interview question is below:

\begin{framed}
\textit{You are working on a summer research project with the condensed matter team investigating the behavior of a novel material with an index of refraction: n (n is a real and positive number) and mass density $\rho$. Another undergraduate student did some calculations about E\&M waves propagating from air (with an index of refraction of 1) to this new material, and obtained the following relationships for the Reflection and Transmitted coefficient (R and T). 
Without going through your friend’s calculations, could you examine the results and see if they make sense?}
\begin{equation*}
R \equiv \frac {I_{R}} {I_{I}}=\frac{\alpha+\beta}{\alpha-\beta} \,\,\,\,\,\,\,\,\,\,\,\,\,\,\,\,\,\,\,\,\,\,\,\,T\equiv \frac {I_{T}} {I_{I}}=\alpha^{2}\beta^{2}(\frac{\rho}{\alpha-\beta})
\end{equation*}
\textit{($I_{I}, I_{R}, I_{T}$ is the intensity of the incoming, reflected and transmitted waves)} \\
\textit{$\alpha$ and $\beta$ were defined as:}\\

$\alpha=\dfrac{\cos{\theta_{T}}}{\cos{\theta_{I}}}$, $\theta_{I}$ \textit{is the incidence angle, and} $\theta_{T}$ \textit{is the transmitted angle.}\\

$\beta=\dfrac{n(new material)}{n(air)}=\dfrac{n}{1}=n$
\end{framed}

Compared to unit analysis, students had more difficulty with checking limits. For example, students had trouble identifying the independent variable (incident angle $\theta_I$). The interviewer had to provide a hint to direct them (3 out of 4) to think about the limits of R and T when $\theta_I$ varies from 0 to 90$^{\circ}$. The expressions of R and T given became negative or infinity when taking extreme limits of $\theta_I$. Only one student noticed the unphysical limits. Physically, we expect the index of refraction to be larger than 1 (index of refraction for vacuum). One student obtained T\textless 0 and didn't note that this is unphysical. Two students talked about taking the lower limit of the index of refraction to 0. In summary, students from the interviews struggled with identifying independent variables,  taking the limits of R and T, and recognizing unphysical limits.

There are many skills invloved when reflecting on a given result, for example, knowing what the physical limits are; being able to vary the limits of one variable and find out the limiting cases, etc. Our interview suggested that students are struggling with the aspect of recognizing the physical limits. The regular exams did not ask students to reflect on their results or any given results, however, they did ask students to do other types of calculation which revealed similar student difficulties. One exam question (see below) gives students Fresnel equations\footnote {Fresnel equations relate the complex amplitudes of the incident, reflected and transmitted waves. \label{footnote3}} for two different cases and asks students which one can have zero reflection. 

\begin{framed}
\textit{In class and then in homework, we derived the following Fresnel equations, assuming linear media with no magnetic effects:}
\textit{E field polarized in the plane of incidence}
\begin{equation*} 
\tilde{E}_{0, r}=\frac{\alpha-\beta}{\alpha+\beta}\tilde{E}_{0, in} \,\,\,\,\,\,\,\,\,\,\,\,\,\,\tilde{E}_{0, t}=\frac{2}{\alpha+\beta}\tilde{E}_{0, in}
\end{equation*}
\textit{E field polarized perpendicular to the plane of incidence}
\begin{equation*}
\tilde {E}_{0,r} = \frac{1-\alpha\beta}{1+\alpha\beta}\; {\tilde E}_{0,in}\,\,\,\,\,\,\,
\tilde{E}_{0,t} = \frac{2}{1+\alpha\beta} {\tilde E}_{0,in}
\end{equation*}
\textit{where }
\begin{displaymath}
\alpha = \frac{\cos{\theta_t}}{\cos{\theta_{in}}} 
\qquad \mbox{and} \qquad
\beta = \frac{n_2}{n_1}
\end{displaymath}
\begin{itemize}
\item[a)]
\textit{Describe a non-trivial physical situation in which the reflected light would have zero intensity.}
\item[b)]
\textit{Describe a non-trivial physical situation in which the electric field vector of the reflected light is 180 degrees out of phase with the E field of the incoming light.}
\item[c)]
\textit{Describe a non-trivial physical situation in which the transmitted light would have zero intensity. (Hint: it is possible, you might have to ``think outside of the box'')}
\end{itemize}

\end{framed}

\noindent Almost a quarter (20\%, 6 out of 30) of the solutions wrote down unphysical or wrong limits such as: the index of refraction becomes negative, zero; velocity of the wave in the media is much larger than the speed of light, $\alpha$ has to be finite, $\alpha$*$\beta$=1 etc. 



\section{\label{sec:tunneling}COMPARISION WITH QUANTUM TUNNELING}
\qr{It is useful to compare student difficulties across subjects to see where there are similarities and/or differences. As mentioned previously}, McKagan \cite{mckagan2008} et al. studied student difficulties with quantum tunneling. A common issue was that students think the energy of the particle is lost after tunneling. One type of reasoning students gave for the energy loss was `` since only part of the wave function is transmitted, only part of the energy is transmitted, with the rest being reflected. " Therefore students think after tunneling, the energy of the particle is lost.  It is true for classical E\&M waves at a boundary that part of the wave is reflected and part of the wave is transmitted. This understanding about classical E\&M waves was also confirmed in our study, in fact, that was a significant aspect of student difficulties when applying BCs (discussed previously in Sec.  \ref{sec:activation} \& \ref{sec:construction}).


It is understandable that students might experience interference of their knowledge and understanding between different contexts. \qr{The word ``inteference'' here refers to a cognitive interference (reported previously in cognitive literature\cite{Wixted2004, Tomlinson2009}), which occurs in learning when there is an interaction between new material and transfer effects of past learned behavior, memories or thoughts that have a negative influence in comprehending new material \cite{Wixted2004, Tomlinson2009}.  Inteference in learning quantum mechanics from classical E\&M waves are understandable because students usually learn the classical waves first and the probability wave idea in quantum is unfamiliar to students (as observed by McKagan.et.al).}  However, interference from quantum mechanics while learning classical E\&M waves is also possible, especially if students are taking these two courses at the same time.  In our study, we observed one example of such interference. When a student was trying to decide if there should be a plus or minus sign between $\vec E_{in}$ and $\vec E_{refl}$, they wrote down: $ 1= Prob (R)+ Prob (T) $ , stating that the total probablity of reflection and transmission should add up to be 1. Then the student continued to write: $ 1-Prob (R)= Prob (T) $, and then confirmed that there should be a minus sign before the $E$ field of the reflected wave. The student conluded that the answer should be:  $\vec E_{in} - \vec E_{refl} = \vec E_{tran}$. When asked by the interviewer what they meant by ``probability", the student said: ``\textit{I'm also taking quantum mechanics and I remember those from my quantum class}".

Another reason students gave for energy loss in quantum tunneling was that they were thinking about the relationship between energy and amplitude (the two are directly related in classical situations) \cite{mckagan2008}. It was pointed out that students' understanding of the energy-amplitude relationship in the classical sense often hinders them from undestanding the probablity waves in the new context (i.e. quantum mechanics). Interestingly, even in the context of classical E\&M waves, students' understanding of the energy-amplitude relationship was also observed to affect their performance with BCs problems.  From both the exam solutions and student interviews, we observed students often wrote down $|{E_{0,in}}|=|{E_{0,refl}}|+|{E_{0,trans}}|$, invoking an idea of energy conservation: part of the wave is reflected and part of the wave is transmitted. Here in classical E\&M context, some students treat energy and wave amplitude interchangable, which hinders them from applying BCs correctly. The idea of conservation of energy was correct, but students replaced energy with the amplitude of the waves and did not activate BCs.

\section{\label{sec:conclusionsanddiscussion}CONCLUSIONS AND DISCUSSIONS}

BCs appear in different contexts throughout in the advanced undergraduate curriculum: classical mechanics, quantum mechanics, and throughout E\&M.  
There has been relatively litte research on student difficulties with BCs. One notable exception is the work by McKagan et al. who reported their findings with student difficulties with quantum mechanics tunneling. They were focusing on the conceptual interpretation of energy conservation during tunneling. 

This paper contributes to the body of research on student difficulties with BCs by presenting an application of the ACER framework to guide analysis of student problem solving using BCs focusing in the context of electromagnetic waves in junior-level electrodynamics. Application of the ACER framework provided an organizing structure for our analysis that helped us identify nodes in students' work where key difficulties appear. It also informed the development of interview protocols that targeted aspects of student problem solving not accessed by traditional exams.

Here we summarize our findings with student difficulties organized by the ACER framework. 

(1) Activation: Our regular exam questions tend to bypass activation by listing BCs explicitly for students and asking them to use BCs. When direct activation is absent (student interviews and question on the CURrENT), many students had trouble activating the correct BCs equations in the context of E\&M waves. They were mostly activating a different type of resource: an idea of conservation of energy, instead of BCs. When shifting to a different context other than E\&M waves (e.g. a CURrENT question about wire with steady current), only a third of the students activated BCs on their own. Most of the interview students couldn't activate BCs even when different cues were provided. When being explicited told to use BCs, some students still didn't think BCs were applicable for this situation.

(2) Construction: Writing complex expressions for E and/or B field of the incoming, reflected and transmitted waves is an important construction element. Common student difficulties with this element include: Finding the correct direction for $\vec {E}_{refl}$; and simplifying the dot product $\vec {k} \cdot \vec{r}$. Other common difficulties with construction include: choosing correct components for the parallel and perpendicular fields; setting up equations only at the boundary, etc. 

(3) Execution: Students were observed to cancel out exponentials terms inappropriately even when the boundary was not located at zero. This execution mistake was not simply a math mistake, since students did not make the same mistake in a pure math context. However, interviews showed that students were still struggling with other aspects of conducting and simplifying complex exponential math.  

(4) Reflection: Our exam questions never asked students to reflect and we see no spontaneous reflection. We constructed interview questions to investigate reflection. Students were having trouble checking the physical limits of the reflected and transmitted coefficient (R\&T) when asked to evaluate the reasonableness of R\&T.
 
These findings have several potential implications for teaching and assessing the use of BCs in electrodynamics and even across subjects. Often in electrodynamics, BCs are strongly associated with the context of E\&M waves and it can be worthwhile to shift the context to non-wave situations (e.g. going back to a static situation).  In our physics class, students often only practice on simple situations (e.g. boundary $z=0$) and this may cause them to over-generalize the results to other situations. It is likely worthwhile to present students a variety of physical situations where there can be both simple and messier results. Students could also benefit from practice on recognizing unphysical limits and taking explicit limits using corresponding independent variables to develop their reflection skills.

It is important to be aware of the limitations to this study. First of all, it only focuses on the use of BCs without exploring the use of complex exponentials by itself, which is also an important element in solving most of these BCs questions. Secondly, another limitation was inherited from the analytical framework (ACER) itself. ACER does not explicitly address the nuance associated with representations. For example, the framework does not explore the translation between verbal, mathematical, graphical, and or pictorial representations, which is almost always required to solve physics problems. \cite{Wilcox2013}. Lastly, while a good number of interviews are conducted, most of the data analyzed are still written solutions. The limitations of written solutions mean that we can observe student difficulties without knowing the reasoning mechanism behind them most of the time.

Our ongoing projects involve analyses of student difficulties with BCs using the theoretical framework of symbolic forms \cite{Sherin1996, Sherin2001}, which will shed light on students' reasoning mechanism. We also plan to expand the investigation of student difficulties across different contexts: in both E\&M and Quantum mechanics.  Further work that we encourage also includes taking the richness of using different representations into consideration. We also plan to further investigate student difficulties on the use of the mathematical tool --- complex exponentials, in order to compare with student difficulties on the use of this physical concept (i.e.\ BCs).


\section{\label{sec:acknowledge}ACKNOWLEDGMENTS}

We gratefully acknowledge contributions of Andreas. Becker, Dimitri Dounas-Frazer and the PER{@}C group. This work was supported by NSF-CCLI grant \#1023208.

\bibliography{BCpaper-draft2}

\begin{thebibliography}{31}%
\makeatletter
\providecommand \@ifxundefined [1]{%
 \@ifx{#1\undefined}
}%
\providecommand \@ifnum [1]{%
 \ifnum #1\expandafter \@firstoftwo
 \else \expandafter \@secondoftwo
 \fi
}%
\providecommand \@ifx [1]{%
 \ifx #1\expandafter \@firstoftwo
 \else \expandafter \@secondoftwo
 \fi
}%
\providecommand \natexlab [1]{#1}%
\providecommand \enquote  [1]{``#1''}%
\providecommand \bibnamefont  [1]{#1}%
\providecommand \bibfnamefont [1]{#1}%
\providecommand \citenamefont [1]{#1}%
\providecommand \href@noop [0]{\@secondoftwo}%
\providecommand \href [0]{\begingroup \@sanitize@url \@href}%
\providecommand \@href[1]{\@@startlink{#1}\@@href}%
\providecommand \@@href[1]{\endgroup#1\@@endlink}%
\providecommand \@sanitize@url [0]{\catcode `\\12\catcode `\$12\catcode
  `\&12\catcode `\#12\catcode `\^12\catcode `\_12\catcode `\%12\relax}%
\providecommand \@@startlink[1]{}%
\providecommand \@@endlink[0]{}%
\providecommand \url  [0]{\begingroup\@sanitize@url \@url }%
\providecommand \@url [1]{\endgroup\@href {#1}{\urlprefix }}%
\providecommand \urlprefix  [0]{URL }%
\providecommand \Eprint [0]{\href }%
\providecommand \doibase [0]{http://dx.doi.org/}%
\providecommand \selectlanguage [0]{\@gobble}%
\providecommand \bibinfo  [0]{\@secondoftwo}%
\providecommand \bibfield  [0]{\@secondoftwo}%
\providecommand \translation [1]{[#1]}%
\providecommand \BibitemOpen [0]{}%
\providecommand \bibitemStop [0]{}%
\providecommand \bibitemNoStop [0]{.\EOS\space}%
\providecommand \EOS [0]{\spacefactor3000\relax}%
\providecommand \BibitemShut  [1]{\csname bibitem#1\endcsname}%
\let\auto@bib@innerbib\@empty
\bibitem [{\citenamefont {Meltzer}\ and\ \citenamefont
  {Thornton}(2012)}]{Meltzer2012}%
  \BibitemOpen
  \bibfield  {author} {\bibinfo {author} {\bibfnamefont {D.~E.}\ \bibnamefont
  {Meltzer}}\ and\ \bibinfo {author} {\bibfnamefont {R.~K.}\ \bibnamefont
  {Thornton}},\ }\href {\doibase http://dx.doi.org/10.1119/1.3678299}
  {\bibfield  {journal} {\bibinfo  {journal} {American Journal of Physics}\
  }\textbf {\bibinfo {volume} {80}},\ \bibinfo {pages} {478} (\bibinfo {year}
  {2012})}\BibitemShut {NoStop}%
\bibitem [{\citenamefont {Caballero}\ \emph {et~al.}(2015)\citenamefont
  {Caballero}, \citenamefont {Wilcox}, \citenamefont {Doughty},\ and\
  \citenamefont {Pollock}}]{caballero2015}%
  \BibitemOpen
  \bibfield  {author} {\bibinfo {author} {\bibfnamefont {M.~D.}\ \bibnamefont
  {Caballero}}, \bibinfo {author} {\bibfnamefont {B.~R.}\ \bibnamefont
  {Wilcox}}, \bibinfo {author} {\bibfnamefont {L.}~\bibnamefont {Doughty}}, \
  and\ \bibinfo {author} {\bibfnamefont {S.~J.}\ \bibnamefont {Pollock}},\
  }\href {http://stacks.iop.org/0143-0807/36/i=6/a=065004} {\bibfield
  {journal} {\bibinfo  {journal} {European Journal of Physics}\ }\textbf
  {\bibinfo {volume} {36}},\ \bibinfo {pages} {065004} (\bibinfo {year}
  {2015})}\BibitemShut {NoStop}%
\bibitem [{\citenamefont {Pepper}\ \emph {et~al.}(2012)\citenamefont {Pepper},
  \citenamefont {Chasteen}, \citenamefont {Pollock},\ and\ \citenamefont
  {Perkins}}]{pepper2012}%
  \BibitemOpen
  \bibfield  {author} {\bibinfo {author} {\bibfnamefont {R.~E.}\ \bibnamefont
  {Pepper}}, \bibinfo {author} {\bibfnamefont {S.~V.}\ \bibnamefont
  {Chasteen}}, \bibinfo {author} {\bibfnamefont {S.~J.}\ \bibnamefont
  {Pollock}}, \ and\ \bibinfo {author} {\bibfnamefont {K.~K.}\ \bibnamefont
  {Perkins}},\ }\href {\doibase 10.1103/PhysRevSTPER.8.010111} {\bibfield
  {journal} {\bibinfo  {journal} {Phys. Rev. ST Phys. Educ. Res.}\ }\textbf
  {\bibinfo {volume} {8}},\ \bibinfo {pages} {010111} (\bibinfo {year}
  {2012})}\BibitemShut {NoStop}%
\bibitem [{\citenamefont {Manogue}\ \emph {et~al.}(2001)\citenamefont
  {Manogue}, \citenamefont {Siemens}, \citenamefont {Tate}, \citenamefont
  {Browne}, \citenamefont {Niess},\ and\ \citenamefont {Wolfer}}]{Manogue2001}%
  \BibitemOpen
  \bibfield  {author} {\bibinfo {author} {\bibfnamefont {C.~A.}\ \bibnamefont
  {Manogue}}, \bibinfo {author} {\bibfnamefont {P.~J.}\ \bibnamefont
  {Siemens}}, \bibinfo {author} {\bibfnamefont {J.}~\bibnamefont {Tate}},
  \bibinfo {author} {\bibfnamefont {K.}~\bibnamefont {Browne}}, \bibinfo
  {author} {\bibfnamefont {M.~L.}\ \bibnamefont {Niess}}, \ and\ \bibinfo
  {author} {\bibfnamefont {A.~J.}\ \bibnamefont {Wolfer}},\ }\href {\doibase
  http://dx.doi.org/10.1119/1.1374248} {\bibfield  {journal} {\bibinfo
  {journal} {American Journal of Physics}\ }\textbf {\bibinfo {volume} {69}},\
  \bibinfo {pages} {978} (\bibinfo {year} {2001})}\BibitemShut {NoStop}%
\bibitem [{\citenamefont {Boas}(1983)}]{mathphysics}%
  \BibitemOpen
  \bibfield  {author} {\bibinfo {author} {\bibfnamefont {M.}~\bibnamefont
  {Boas}},\ }\href@noop {} {\emph {\bibinfo {title} {Mathematical methods in
  the physical sciences, 2nd. Edition}}}\ (\bibinfo {address} {Wiley},\
  \bibinfo {year} {1983})\BibitemShut {NoStop}%
\bibitem [{\citenamefont {Gladwell}(2008)}]{Gladwell2008}%
  \BibitemOpen
  \bibfield  {author} {\bibinfo {author} {\bibfnamefont {I.}~\bibnamefont
  {Gladwell}},\ }\href@noop {} {\bibfield  {journal} {\bibinfo  {journal}
  {Scholarpedia}\ }\textbf {\bibinfo {volume} {3}},\ \bibinfo {pages} {2853}
  (\bibinfo {year} {2008})},\ \bibinfo {note} {revision number
  91077}\BibitemShut {NoStop}%
\bibitem [{\citenamefont {Taylor}(2005)}]{classicalmechanics}%
  \BibitemOpen
  \bibfield  {author} {\bibinfo {author} {\bibfnamefont {J.}~\bibnamefont
  {Taylor}},\ }\href@noop {} {\emph {\bibinfo {title} {Classical mechanics}}}\
  (\bibinfo {address} {University Science Books},\ \bibinfo {year}
  {2005})\BibitemShut {NoStop}%
\bibitem [{\citenamefont {Griffiths}()}]{Griffiths}%
  \BibitemOpen
  \bibfield  {author} {\bibinfo {author} {\bibfnamefont {D.}~\bibnamefont
  {Griffiths}},\ }\href@noop {} {\emph {\bibinfo {title} {{Introduction to
  Electrodynamics, 3rd Ed. }}}}\ (\bibinfo {address} {{Upper-Saddle River
  NJ}})\BibitemShut {NoStop}%
\bibitem [{\citenamefont {de~la Madrid}(2003)}]{madrid2003}%
  \BibitemOpen
  \bibfield  {author} {\bibinfo {author} {\bibfnamefont {R.}~\bibnamefont
  {de~la Madrid}},\ }\href {arXiv:quant-ph/0302184} {\bibfield  {journal}
  {\bibinfo  {journal} {arXiv:quant-ph/0302184}\ } (\bibinfo {year}
  {2003})}\BibitemShut {NoStop}%
\bibitem [{\citenamefont {Wilcox}\ and\ \citenamefont
  {Pollock}(2015{\natexlab{a}})}]{Wilcox2015}%
  \BibitemOpen
  \bibfield  {author} {\bibinfo {author} {\bibfnamefont {B.~R.}\ \bibnamefont
  {Wilcox}}\ and\ \bibinfo {author} {\bibfnamefont {S.~J.}\ \bibnamefont
  {Pollock}},\ }\href {\doibase 10.1103/PhysRevSTPER.11.020131} {\bibfield
  {journal} {\bibinfo  {journal} {Phys. Rev. ST Phys. Educ. Res.}\ }\textbf
  {\bibinfo {volume} {11}},\ \bibinfo {pages} {020131} (\bibinfo {year}
  {2015}{\natexlab{a}})}\BibitemShut {NoStop}%
\bibitem [{\citenamefont {McKagan}\ \emph {et~al.}(2008)\citenamefont
  {McKagan}, \citenamefont {Perkins},\ and\ \citenamefont
  {Wieman}}]{mckagan2008}%
  \BibitemOpen
  \bibfield  {author} {\bibinfo {author} {\bibfnamefont {S.~B.}\ \bibnamefont
  {McKagan}}, \bibinfo {author} {\bibfnamefont {K.~K.}\ \bibnamefont
  {Perkins}}, \ and\ \bibinfo {author} {\bibfnamefont {C.~E.}\ \bibnamefont
  {Wieman}},\ }\href {\doibase 10.1103/PhysRevSTPER.4.020103} {\bibfield
  {journal} {\bibinfo  {journal} {Phys. Rev. ST Phys. Educ. Res.}\ }\textbf
  {\bibinfo {volume} {4}},\ \bibinfo {pages} {020103} (\bibinfo {year}
  {2008})}\BibitemShut {NoStop}%
\bibitem [{\citenamefont {Wilcox}\ \emph {et~al.}(2013)\citenamefont {Wilcox},
  \citenamefont {Caballero}, \citenamefont {Rehn},\ and\ \citenamefont
  {Pollock}}]{Wilcox2013}%
  \BibitemOpen
  \bibfield  {author} {\bibinfo {author} {\bibfnamefont {B.~R.}\ \bibnamefont
  {Wilcox}}, \bibinfo {author} {\bibfnamefont {M.~D.}\ \bibnamefont
  {Caballero}}, \bibinfo {author} {\bibfnamefont {D.~A.}\ \bibnamefont {Rehn}},
  \ and\ \bibinfo {author} {\bibfnamefont {S.~J.}\ \bibnamefont {Pollock}},\
  }\href {\doibase 10.1103/PhysRevSTPER.9.020119} {\bibfield  {journal}
  {\bibinfo  {journal} {Phys. Rev. ST Phys. Educ. Res.}\ }\textbf {\bibinfo
  {volume} {9}},\ \bibinfo {pages} {020119} (\bibinfo {year}
  {2013})}\BibitemShut {NoStop}%
\bibitem [{\citenamefont {Wilcox}\ and\ \citenamefont
  {Pollock}(2015{\natexlab{b}})}]{Wilcoxdeltafunction}%
  \BibitemOpen
  \bibfield  {author} {\bibinfo {author} {\bibfnamefont {B.~R.}\ \bibnamefont
  {Wilcox}}\ and\ \bibinfo {author} {\bibfnamefont {S.~J.}\ \bibnamefont
  {Pollock}},\ }\href {\doibase 10.1103/PhysRevSTPER.11.010108} {\bibfield
  {journal} {\bibinfo  {journal} {Phys. Rev. ST Phys. Educ. Res.}\ }\textbf
  {\bibinfo {volume} {11}},\ \bibinfo {pages} {010108} (\bibinfo {year}
  {2015}{\natexlab{b}})}\BibitemShut {NoStop}%
\bibitem [{\citenamefont {Wilcox}\ and\ \citenamefont
  {Pollock}(2014)}]{Wilcox2014}%
  \BibitemOpen
  \bibfield  {author} {\bibinfo {author} {\bibfnamefont {B.}~\bibnamefont
  {Wilcox}}\ and\ \bibinfo {author} {\bibfnamefont {S.~J.}\ \bibnamefont
  {Pollock}},\ }in\ \href@noop {} {\emph {\bibinfo {booktitle} {Physics
  Education Research Conference 2014}}},\ \bibinfo {series and number} {PER
  Conference}\ (\bibinfo {address} {Minneapolis, MN},\ \bibinfo {year} {2014})\
  pp.\ \bibinfo {pages} {271--274}\BibitemShut {NoStop}%
\bibitem [{\citenamefont {Hammer}(2000)}]{Hammer2000}%
  \BibitemOpen
  \bibfield  {author} {\bibinfo {author} {\bibfnamefont {D.}~\bibnamefont
  {Hammer}},\ }\href {\doibase http://dx.doi.org/10.1119/1.19520} {\bibfield
  {journal} {\bibinfo  {journal} {American Journal of Physics}\ }\textbf
  {\bibinfo {volume} {68}},\ \bibinfo {pages} {S52} (\bibinfo {year}
  {2000})}\BibitemShut {NoStop}%
\bibitem [{\citenamefont {Tuminaro}(2004)}]{Tuminaro}%
  \BibitemOpen
  \bibfield  {author} {\bibinfo {author} {\bibfnamefont {J.}~\bibnamefont
  {Tuminaro}},\ }\href@noop {} {\bibfield  {journal} {\bibinfo  {journal}
  {Ph.D. dissertation, University of Maryland}\ } (\bibinfo {year}
  {2004})}\BibitemShut {NoStop}%
\bibitem [{\citenamefont {Tuminaro}\ and\ \citenamefont
  {Redish}(2007)}]{Tuminaro2007}%
  \BibitemOpen
  \bibfield  {author} {\bibinfo {author} {\bibfnamefont {J.}~\bibnamefont
  {Tuminaro}}\ and\ \bibinfo {author} {\bibfnamefont {E.~F.}\ \bibnamefont
  {Redish}},\ }\href {\doibase 10.1103/PhysRevSTPER.3.020101} {\bibfield
  {journal} {\bibinfo  {journal} {Phys. Rev. ST Phys. Educ. Res.}\ }\textbf
  {\bibinfo {volume} {3}},\ \bibinfo {pages} {020101} (\bibinfo {year}
  {2007})}\BibitemShut {NoStop}%
\bibitem [{\citenamefont {Tuminaro}(2008)}]{Bing}%
  \BibitemOpen
  \bibfield  {author} {\bibinfo {author} {\bibfnamefont {B.}~\bibnamefont
  {Tuminaro}},\ }\href@noop {} {\bibfield  {journal} {\bibinfo  {journal}
  {Ph.D. dissertation, University of Maryland}\ } (\bibinfo {year}
  {2008})}\BibitemShut {NoStop}%
\bibitem [{\citenamefont {diSessa}(1993)}]{diSessa1993}%
  \BibitemOpen
  \bibfield  {author} {\bibinfo {author} {\bibfnamefont {A.}~\bibnamefont
  {diSessa}},\ }\href@noop {} {\bibfield  {journal} {\bibinfo  {journal} {Cogn.
  Instruct.}\ }\textbf {\bibinfo {volume} {10}},\ \bibinfo {pages} {105}
  (\bibinfo {year} {1993})}\BibitemShut {NoStop}%
\bibitem [{\citenamefont {diSessa}\ and\ \citenamefont
  {Sherin}(1998)}]{diSessa1998}%
  \BibitemOpen
  \bibfield  {author} {\bibinfo {author} {\bibfnamefont {A.}~\bibnamefont
  {diSessa}}\ and\ \bibinfo {author} {\bibfnamefont {B.}~\bibnamefont
  {Sherin}},\ }\href@noop {} {\bibfield  {journal} {\bibinfo  {journal} {Int.J.
  Sci. Educ.}\ }\textbf {\bibinfo {volume} {20}},\ \bibinfo {pages} {1155}
  (\bibinfo {year} {1998})}\BibitemShut {NoStop}%
\bibitem [{\citenamefont {Sherin}(2008)}]{Sherin1996}%
  \BibitemOpen
  \bibfield  {author} {\bibinfo {author} {\bibfnamefont {B.}~\bibnamefont
  {Sherin}},\ }\href@noop {} {\bibfield  {journal} {\bibinfo  {journal} {Ph.D.
  dissertation, University of California, Berkeley}\ } (\bibinfo {year}
  {2008})}\BibitemShut {NoStop}%
\bibitem [{\citenamefont {diSessa}\ and\ \citenamefont
  {Sherin}(2001)}]{Sherin2001}%
  \BibitemOpen
  \bibfield  {author} {\bibinfo {author} {\bibfnamefont {A.}~\bibnamefont
  {diSessa}}\ and\ \bibinfo {author} {\bibfnamefont {B.}~\bibnamefont
  {Sherin}},\ }\href@noop {} {\bibfield  {journal} {\bibinfo  {journal} {Cogn.
  Instruct.}\ }\textbf {\bibinfo {volume} {19}},\ \bibinfo {pages} {479}
  (\bibinfo {year} {2001})}\BibitemShut {NoStop}%
\bibitem [{\citenamefont {Collins}\ and\ \citenamefont
  {Ferguson}(1993)}]{Collins1993}%
  \BibitemOpen
  \bibfield  {author} {\bibinfo {author} {\bibfnamefont {A.}~\bibnamefont
  {Collins}}\ and\ \bibinfo {author} {\bibfnamefont {W.}~\bibnamefont
  {Ferguson}},\ }\href@noop {} {\bibfield  {journal} {\bibinfo  {journal}
  {Educ. Psychol.}\ }\textbf {\bibinfo {volume} {28}},\ \bibinfo {pages} {25}
  (\bibinfo {year} {1993})}\BibitemShut {NoStop}%
\bibitem [{\citenamefont {Wilcox}(2015)}]{wilcox2015thesis}%
  \BibitemOpen
  \bibfield  {author} {\bibinfo {author} {\bibfnamefont {B.~R.}\ \bibnamefont
  {Wilcox}},\ }\href@noop {} {\enquote {\bibinfo {title} {New tools for
  investigating student learning in upper-division electrostatics},}\ }
  (\bibinfo {year} {2015})\BibitemShut {NoStop}%
\bibitem [{\citenamefont {Chasteen}\ \emph {et~al.}(2012)\citenamefont
  {Chasteen}, \citenamefont {Pollock}, \citenamefont {Pepper},\ and\
  \citenamefont {Perkins}}]{Chasteen2012}%
  \BibitemOpen
  \bibfield  {author} {\bibinfo {author} {\bibfnamefont {S.~V.}\ \bibnamefont
  {Chasteen}}, \bibinfo {author} {\bibfnamefont {S.~J.}\ \bibnamefont
  {Pollock}}, \bibinfo {author} {\bibfnamefont {R.~E.}\ \bibnamefont {Pepper}},
  \ and\ \bibinfo {author} {\bibfnamefont {K.~K.}\ \bibnamefont {Perkins}},\
  }\href {\doibase 10.1103/PhysRevSTPER.8.020107} {\bibfield  {journal}
  {\bibinfo  {journal} {Phys. Rev. ST Phys. Educ. Res.}\ }\textbf {\bibinfo
  {volume} {8}},\ \bibinfo {pages} {020107} (\bibinfo {year}
  {2012})}\BibitemShut {NoStop}%
\bibitem [{\citenamefont {Mazur}(1997)}]{Mazur}%
  \BibitemOpen
  \bibfield  {author} {\bibinfo {author} {\bibfnamefont {E.}~\bibnamefont
  {Mazur}},\ }\href@noop {} {\emph {\bibinfo {title} {Peer Instruction:
  User’s Manual, Series in Educational Innovation}}}\ (\bibinfo {address}
  {Prentice Hall, Upper Saddle River,NJ},\ \bibinfo {year} {1997})\BibitemShut
  {NoStop}%
\bibitem [{\citenamefont {Baily}\ \emph {et~al.}(2012)\citenamefont {Baily},
  \citenamefont {Dubson},\ and\ \citenamefont {Pollock}}]{Baily2012}%
  \BibitemOpen
  \bibfield  {author} {\bibinfo {author} {\bibfnamefont {C.}~\bibnamefont
  {Baily}}, \bibinfo {author} {\bibfnamefont {M.}~\bibnamefont {Dubson}}, \
  and\ \bibinfo {author} {\bibfnamefont {S.~J.}\ \bibnamefont {Pollock}},\ }in\
  \href@noop {} {\emph {\bibinfo {booktitle} {Physics Education Research
  Conference 2012}}},\ \bibinfo {series and number} {PER Conference}\ (\bibinfo
  {address} {Philadelphia, PA},\ \bibinfo {year} {2012})\ pp.\ \bibinfo {pages}
  {54--57}\BibitemShut {NoStop}%
\bibitem [{\citenamefont {Ryan}\ \emph {et~al.}(2014)\citenamefont {Ryan},
  \citenamefont {Astolfi}, \citenamefont {Baily},\ and\ \citenamefont
  {Pollock}}]{Ryan2014}%
  \BibitemOpen
  \bibfield  {author} {\bibinfo {author} {\bibfnamefont {Q.~X.}\ \bibnamefont
  {Ryan}}, \bibinfo {author} {\bibfnamefont {C.}~\bibnamefont {Astolfi}},
  \bibinfo {author} {\bibfnamefont {C.}~\bibnamefont {Baily}}, \ and\ \bibinfo
  {author} {\bibfnamefont {S.~J.}\ \bibnamefont {Pollock}},\ }in\ \href@noop {}
  {\emph {\bibinfo {booktitle} {Physics Education Research Conference 2014}}},\
  \bibinfo {series and number} {PER Conference}\ (\bibinfo {address}
  {Minneapolis, MN},\ \bibinfo {year} {2014})\ pp.\ \bibinfo {pages}
  {231--234}\BibitemShut {NoStop}%
\bibitem [{per()}]{perwebsite}%
  \BibitemOpen
  \href {http://www.colorado.edu/physics/EducationIssues/Electrodynamics/}
  {}\bibinfo {howpublished}
  {\url{http://www.colorado.edu/physics/EducationIssues/Electrodynamics/}}\BibitemShut
  {NoStop}%
\bibitem [{\citenamefont {Wixted}(2004)}]{Wixted2004}%
  \BibitemOpen
  \bibfield  {author} {\bibinfo {author} {\bibfnamefont {J.~T.}\ \bibnamefont
  {Wixted}},\ }\href {\doibase 10.1146/annurev.psych.55.090902.141555}
  {\bibfield  {journal} {\bibinfo  {journal} {Annual Review of Psychology}\
  }\textbf {\bibinfo {volume} {55}},\ \bibinfo {pages} {235} (\bibinfo {year}
  {2004})},\ \bibinfo {note} {pMID: 14744216},\ \Eprint
  {http://arxiv.org/abs/http://dx.doi.org/10.1146/annurev.psych.55.090902.141555}
  {http://dx.doi.org/10.1146/annurev.psych.55.090902.141555} \BibitemShut
  {NoStop}%
\bibitem [{\citenamefont {Tomlinson}\ \emph {et~al.}(2009)\citenamefont
  {Tomlinson}, \citenamefont {Huber}, \citenamefont {Rieth},\ and\
  \citenamefont {Davelaar}}]{Tomlinson2009}%
  \BibitemOpen
  \bibfield  {author} {\bibinfo {author} {\bibfnamefont {T.~D.}\ \bibnamefont
  {Tomlinson}}, \bibinfo {author} {\bibfnamefont {D.~E.}\ \bibnamefont
  {Huber}}, \bibinfo {author} {\bibfnamefont {C.~A.}\ \bibnamefont {Rieth}}, \
  and\ \bibinfo {author} {\bibfnamefont {E.~J.}\ \bibnamefont {Davelaar}},\
  }\href {\doibase 10.1073/pnas.0813370106} {\bibfield  {journal} {\bibinfo
  {journal} {Proceedings of the National Academy of Sciences}\ }\textbf
  {\bibinfo {volume} {106}},\ \bibinfo {pages} {15588} (\bibinfo {year}
  {2009})},\ \Eprint
  {http://arxiv.org/abs/http://www.pnas.org/content/106/37/15588.full.pdf}
  {http://www.pnas.org/content/106/37/15588.full.pdf} \BibitemShut {NoStop}%
\end{thebibliography}%

\end{document}